\documentclass[fleqn,usenatbib]{mnras}

\usepackage{newtxtext,newtxmath}

\usepackage[T1]{fontenc}

\DeclareRobustCommand{\VAN}[3]{#2}
\let\VANthebibliography\thebibliography
\def\thebibliography{\DeclareRobustCommand{\VAN}[3]{##3}\VANthebibliography}

\usepackage{graphicx}	
\usepackage{amssymb,amsmath,latexsym,mathrsfs}

\usepackage{physics}
\usepackage{mathtools}
\usepackage{multirow}
\usepackage{booktabs}
\usepackage{adjustbox}
\usepackage{pdflscape}
\usepackage[shortlabels]{enumitem}
\usepackage{bm}
\usepackage[usenames,dvipsnames]{color}
\usepackage{makecell}
\usepackage{ulem}  
\usepackage[dvipsnames]{xcolor}

\newcommand{\bfalpha}{{\boldsymbol{\alpha}}}
\newcommand{\bftheta}{{\boldsymbol{\theta}}}

\newcommand{\bfx}{{\boldsymbol{x}}}
\newcommand{\bfz}{{\boldsymbol{z}}}

\newcommand{\bfk}{{\boldsymbol{k}}}

\newcommand{\boldmu}{\boldsymbol{\mu}}

\title[Gaussian NPE for Cosmological ICs]{Fast Sampling of Cosmological Initial Conditions with Gaussian Neural Posterior Estimation}

\author[Oleg Savchenko et al.]{
Oleg Savchenko,$^{1}$\thanks{E-mail: o.savchenko@uva.nl}
Guillermo Franco Abell\'{a}n,$^{1}$
Florian List,$^{2}$
Noemi Anau Montel,$^{1}$
and Christoph Weniger$^{1}$
\\
$^{1}$GRAPPA Institute, Institute for Theoretical Physics Amsterdam, University of Amsterdam, Science Park 904, 1098 XH Amsterdam, The Netherlands\\
$^{2}$Department of Astrophysics, University of Vienna, Türkenschanzstraße 17, 1180 Vienna, Austria\\
}

\date{Accepted XXX. Received YYY; in original form ZZZ}

\pubyear{\the\year{}}

\begin{document}
\label{firstpage}
\pagerange{\pageref{firstpage}--\pageref{lastpage}}
\maketitle

\begin{abstract}
Knowledge of the primordial matter density field from which the large-scale structure of the Universe emerged over cosmic time is of fundamental importance for cosmology. However, reconstructing these cosmological initial conditions from late-time observations is a notoriously difficult task, which requires advanced cosmological simulators and sophisticated statistical methods to explore a multi-million-dimensional parameter space. We show how simulation-based inference (SBI) can be used to tackle this problem and to obtain data-constrained realisations of the primordial dark matter density field in a simulation-efficient way with general non-differentiable simulators. Our method is applicable to full high-resolution dark matter N-body simulations and is based on modelling the posterior distribution of the constrained initial conditions to be Gaussian with a diagonal covariance matrix in Fourier space. As a result, we can generate thousands of posterior samples within seconds on a single GPU, orders of magnitude faster than existing methods, paving the way for sequential SBI for cosmological fields. Furthermore, we perform an analytical fit of the estimated dependence of the covariance on the wavenumber, effectively transforming any point-estimator of initial conditions into a fast sampler. We test the validity of our obtained samples by comparing them to the true values with summary statistics and performing a Bayesian consistency test.
\end{abstract}




\section{Introduction} \label{sec:intro}

The current era of precise and abundant cosmological data has firmly established the standard model of cosmology, which shows remarkable agreement with a wide variety of observations and is now constrained to percent-level precision (see, e.g., \citealt{Peebles:2024txt, Efstathiou:2024dvn}). One of the major probes of this model, which still holds significant untapped potential, is the three-dimensional cosmic web distribution of galaxies, referred to as the large-scale structure (LSS) of the Universe. Next-generation galaxy surveys, such as the Dark Energy Spectroscopic Instrument (DESI, \citealt{Levi19}), \textit{Euclid} \citep{Laureijs11}, the Nancy Grace Roman Space Telescope \citep{Spergel15}, and the Vera Rubin Observatory \citep{Ivezic19}, will provide an enormous wealth of new information about the LSS, covering scales from a few Mpc up to the size of the cosmological horizon.  

However, analysing these vast data volumes often requires computationally expensive simulations and exploration of high-dimensional parameter spaces, posing a serious challenge for traditional statistical inference techniques and, in many cases, rendering them effectively infeasible. The necessity to go beyond standard methods, combined with recent revolutionary progress in deep learning, has led to the emergence of powerful alternative approaches capable of tackling these pressing challenges \citep{2023PASA...40....1H}. A promising avenue of research in this direction is simulation-based inference (SBI, \citealt{Cranmer_2020}). All SBI algorithms perform inference using a generative model of the data, i.e. a stochastic simulator that maps model parameters to data realisations, thus effectively sampling an (implicit) likelihood. SBI has already been successfully applied to a variety of low-dimensional tasks in astrophysics and cosmology ~\cite[e.g.,][]{Makinen:2021nly, Cole:2021gwr, Mishra_Sharma_2022, Montel:2022fhv, Karchev:2022xyn, SimBIG:2023ywd, DES:2023qwe, FrancoAbellan:2024tbj, vonWietersheim-Kramsta:2024cks}), but applying it to high-dimensional problems --- such as image or 3D field reconstruction, which often involves inference of $\mathrm{\gtrsim} \, 10^6$ parameters --- still remains an area of active ongoing research ~\cite[e.g.,][]{List:2023jwo, Floss:2024gqk, Sabti:2024jff}.

\begin{table*}
	\centering
	\caption{Overview of selected 3D cosmological matter field reconstruction methods from the literature, alongside our work (this table is not exhaustive but highlights some notable works for comparison). We compare the sizes and grid resolutions of the reconstructed simulation boxes, the implemented methods, and whether they allow sampling the posterior or only getting point estimates, together with the computational resources required for convergence or training}.
	\label{tab:comparison}
	\begin{tabular}{lccccc}
		\hline
		& \makecell{Box size, \\ $\mathrm{Gpc} / h$} & \makecell{Grid \\ resolution} & \makecell{Reconstruction \\ method} & \makecell{Posterior \\ sampling} & Computational resource \\
		\hline
            \texttt{BORG} (\citealt{Jasche:2018oym}) & 0.6777 & $256^3$ & HMC & \textcolor{green}{\checkmark}  & $\sim 10^6$ CPU hours for $\sim 10^3$ samples \\
            \citet{Schmittfull_2017} & 0.5 & $512^3$ & Iterative & \textcolor{red}{$\times$} & 3 CPU hours \\
            \citet{Feng_2018} & 0.4 & $128^3$ & \makecell{Gradient-based MAP \\ estimation} & \textcolor{red}{$\times$} & N/A \\
            \citet{modi2021cosmicrim} & 0.4 & $64^3$ & \makecell{Recurrent inference \\ machines} & \textcolor{red}{$\times$}  & N/A \\
            \citet{Shallue:2022mhf} & $\simeq 0.44$ & $128^3$ & CNN & \textcolor{red}{$\times$} & \makecell{1 \texttt{Nvidia GeForce GTX 1080 Ti} GPU,\\4-12 hours} \\
		\citet{Jindal:2023qew} & 0.25 & $128^3$ & V-Net & \textcolor{red}{$\times$} & N/A \\
            \citet{Legin:2023jxc} & 1 & $128^3$ & Score-based diffusion & \textcolor{green}{\checkmark} & 4 \texttt{Nvidia A100} 80GB GPUs, 24 hours \\
            \citet{Floss:2023ylq} & 1 & $256^3$ & U-Net & \textcolor{red}{$\times$} & 4 \texttt{Nvidia A100} 40GB GPUs, 4 hours \\
            \citet{Wang:2023hgm} & 0.5 & $256^3$ & U-Net & \textcolor{red}{$\times$}  & N/A \\
            \citet{Bayer:2023rmj} & 0.256 & $64^3$ & \makecell{Microcanonical Langevin \\ Monte Carlo} & \textcolor{green}{\checkmark} & N/A \\
            \citet{Doeser:2023yzv} & 0.25 & $128^3$ & \makecell{HMC + differentiable \\ emulator} & \textcolor{green}{\checkmark} & N/A \\
            \citet{Sabti:2024jff} & $\approx 0.5$ & $128^3$ & Stochastic interpolants & \textcolor{green}{\checkmark} & 1 \texttt{Nvidia A100} 80GB GPU, 3 days \\
            This work & 1 & $128^3$ & Gaussian NPE & \textcolor{green}{\checkmark} & 1 \texttt{Nvidia A100} 40GB GPU, 1.5 hour \\
		\hline
	\end{tabular}
\end{table*}

A prominent example of a high-dimensional inference task in the cosmological setting is the problem of reconstructing the cosmological initial conditions (ICs) from which the observed LSS distribution emerged. While the present-day cosmic web has a highly complex, non-linear and non-Gaussian nature, it originates from a much simpler state that is well characterised by linear Gaussian fluctuations \citep{2020moco.book.....D}. With data-constrained realisations of these ICs at hand, one can then forward-simulate their evolution in time, and reconstruct the local matter distribution, as well as the velocity fields and structure formation histories. This has already found important physical applications, for instance, in the context of primordial magnetic fields \citep{Hutschenreuter:2018vkr}, de-biasing $H_0$ inference from standard sirens \citep{Mukherjee:2019qmm}, or the cosmic neutrino background \citep{Elbers:2023mdr}. Most of the standard cosmological analyses rely on using the matter two-point function, which saturates the information content of the LSS only on relatively large scales where the Universe still evolves linearly. In contrast, most of the new information content delivered by the upcoming surveys will come from the small scales where the Universe becomes highly non-linear and non-Gaussian \citep{Beyond-2pt:2024mqz}. Accessing the full field-level information allows extracting all the information available in higher-order statistics, thereby significantly improving constraints on cosmological parameters ~\cite[e.g.,][]{Leclercq:2021ctr, Makinen:2021nly, Andrews:2022nvv, Zhou:2023fey, SimBIG:2023ywd, Floss:2023ylq, Nguyen:2024yth, Euclid:2024ris}.

Early attempts to reconstruct the cosmological ICs using methods such as the least-action principle or optimal mass transportation date back several decades \citep{1989ApJ...344L..53P, 2002Natur.417..260F, Brenier:2003xs}. However, a unique recovery of ICs from a late-time matter density field is generally impossible because the early-to-late time state mapping becomes non-injective in the post-shell-crossing (multi-stream) regime, meaning that different ICs can lead to the same late-time density field \citep{Crocce:2005xz}. Moreover, on halo scales that have decoupled from the Hubble flow, structure formation becomes highly chaotic, which makes trivially inverting the arrow of time in the equations of motion infeasible due to the amplification of any measurement noise (further compounded by the lack of full phase-space information). Even on linear scales, backward integration of the equations of motion is complicated by spurious decaying modes arising from noise or numerical errors, which will grow as $a \to 0$ \citep{1992ApJ...391..443N}. Consequently, analytical descriptions that only account for growing modes ~\cite[such as the Zel'dovich approximation,][]{1970A&A.....5...84Z} are often employed to reconstruct the ICs ~\cite[e.g.,][]{Brenier:2003xs}.

Given this ill-posedness of the problem, probabilistic (Bayesian) reconstruction techniques come as a natural tool for the analysis. This approach was pioneered in the Bayesian Origin Reconstruction from Galaxies (\texttt{BORG}) project (\citealt{Jasche:2012kq, Jasche:2018oym}). \texttt{BORG} uses a Hamiltonian Monte Carlo (HMC) sampling algorithm to efficiently explore the high-dimensional parameter space by leveraging the gradient of the likelihood with respect to the cosmological ICs. While still being the state-of-the-art approach, \texttt{BORG} requires differentiable simulators (which often rely on approximation schemes) and explicitly defined likelihood forms. In addition, the expensive HMC procedure requires months of computation time on clusters for the algorithm to converge, producing a chain of correlated samples that are specific to the given setup and cannot easily be reused. Despite these challenges, impressive results have been obtained with \texttt{BORG}, as well as other sampling approaches based on exploiting the likelihood gradient information ~\cite[e.g.,][]{Kostic:2022vok, Bayer:2023rmj}, highlighting the value of field-level inference.

With the recent advent of machine learning methods in scientific data analysis, there have been a number of works which apply those novel techniques to the problem of field reconstruction in astrophysics and cosmology, and specifically to the problem of cosmological IC reconstruction (see Tab. \ref{tab:comparison} for a comparison between different approaches in the literature and this work). Most of them rely on training a point estimator (usually a U-Net type neural network) to predict one field conditioned on the observation of another field ~\cite[e.g.,][]{Schmittfull_2017, Hada:2018fde, Feng_2018, modi2020flowpm, Hong:2020gvx, modi2021cosmicrim, Shallue:2022mhf, Bayer:2022vid, Jindal:2023qew, Floss:2023ylq, Qin:2023dew, Wang:2023hgm, Chen:2023uup, Horowitz:2023ejo, Shi:2025zoz}. While often being able to obtain very accurate ICs that are consistent with the ground truth, point estimators only yield a `best-fit' estimate without quantifying the uncertainty of the achieved reconstruction. Recently, score-based generative models have shown promising results in modelling the conditional distributions of cosmological fields ~\cite[e.g.,][]{Legin:2023jxc, Park:2023ync, Bourdin:2024zyj}. These methods train neural networks to approximate the gradient of the negative log posterior distribution with respect to the model parameters (the score), and then use these trained networks to generate samples via a reverse-diffusion process. While achieving impressive results, these kind of models are still quite computationally demanding -- even at inference time -- and often lack interpretability. Another promising direction of research is the use of field-level emulators to accelerate modelling of the non-linear gravitational structure formation ~\cite[e.g.,][]{Jamieson:2022lqc, Jamieson:2024fsp}, which e.g. have been recently integrated into the \texttt{BORG} pipeline \citep{Doeser:2023yzv}. 

In this work, we propose a simple and fast method to generate samples of the initial matter density field conditioned on the observation of the present-day matter density field. Our method models the likelihood as Gaussian with respect to the initial density field values (i.e.\ the parameters to be inferred, \textit{not} the observation), with the precision matrix being diagonal in Fourier space. Since the prior is also Gaussian, this results in a Gaussian form of the posterior, with the trainable parts of the model consisting of the U-Net--based maximum a posteriori (MAP) estimator and the diagonal values of the likelihood precision matrix. The simplicity of the method makes it extremely fast both in terms of training ($\sim 1.5 \, \mathrm{h}$ on a single GPU at $128^3$ resolution) and sampling ($\lesssim 3 \, \mathrm{s}$ for $1000$ samples). Crucially, as the forward model does not need to be differentiable, this method can employ full $N$-body simulations without relying on approximation schemes. We also provide an approximate analytical fit of the obtained precision matrix as a function of wavenumber $k$, enabling uncertainty estimation for field reconstruction, thereby turning any IC point estimator into a fast sampler.

The paper is structured as follows. We provide the general description of the method in Section~\ref{sec:methodology}. In Section~\ref{sec:results}, we show how the trained model allows generating ICs samples, and compare them to the ground truth with summary statistics and a Bayesian consistency test. Moreover, we study the structure of the obtained posterior precision matrix and draw its implications. We conclude in Section~\ref{sec:conclusion} with the discussion of our results and an outlook to future developments.

\begin{figure*}
	\includegraphics[width=16cm]{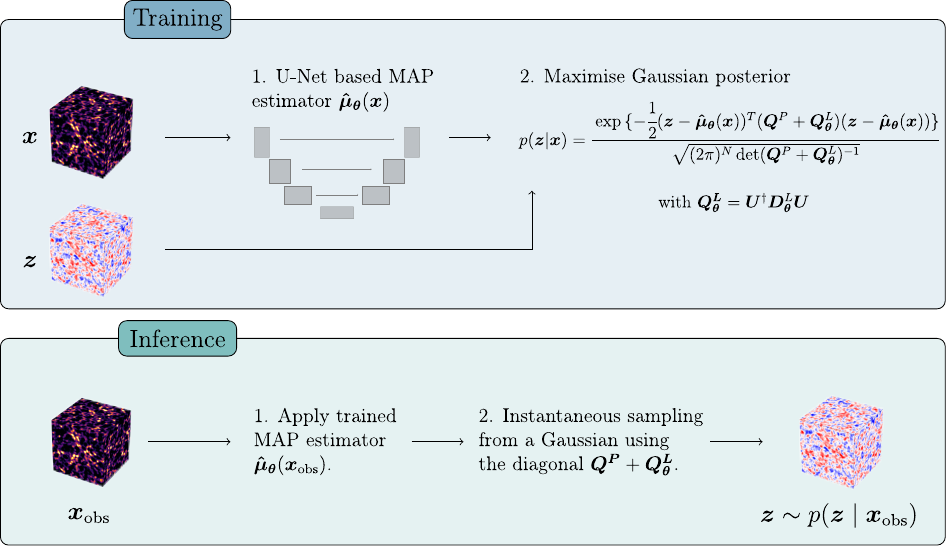}
    \caption{Schematic description of the method presented in this work. The posterior probability distribution of initial conditions $\bfz$ for a given observation of the final conditions $\bfx_{\rm obs}$ is modelled to be Gaussian in $\bfz$, with the likelihood precision matrix $\bm Q_\bftheta^L$ being diagonal in Fourier space. This leads to the ansatz for the posterior (\ref{eq:posterior}). The MAP estimator $\mathbf{\hat{\boldmu}}_\bftheta$ and the likelihood precision matrix $\bm Q_\bftheta^L$ are then simultaneously trained to maximise the posterior probability for the training set under the assumed probability distribution. After training, which takes $\sim 1.5 \, \mathrm{h}$ on a single GPU at $128^3$ resolution, we can obtain samples of the initial conditions in an almost instantaneous way given the Gaussian approximation ($\lesssim 3 \, \mathrm{s}$ for $1000$ samples).}
    \label{fig:diagram_method}
\end{figure*}

\section{Methodology} \label{sec:methodology}
Here, we introduce our Gaussian neural posterior estimation (NPE) framework to reconstruct cosmological initial conditions from late-time density fields, schematically depicted in Fig.~\ref{fig:diagram_method}. We first describe the simulation data used for training and validation (Section~\ref{subsec:simulation_data}). Next, we formalise the high-dimensional statistical problem of mapping a final density field to its initial conditions, motivating our choice of a Gaussian likelihood ansatz (Section~\ref{subsec:statistical_problem}). Finally, we detail our training strategy, outlining the loss function and practical considerations for optimisation (Section~\ref{subsec:training}).

\subsection{Simulation data} \label{subsec:simulation_data}

In contrast to reconstruction methods that rely on Hamiltonian Monte Carlo sampling and thus need the gradient of the likelihood, our method does not require the simulator to be differentiable. This, together with the fact that a relatively small training set is sufficient, enables the use of full state-of-the-art cosmological $N$-body codes as forward models for our reconstruction method.

Our training/validation dataset consists of 2000 pairs of initial (at redshift $z = 127$) and present-day ($z = 0$) dark matter overdensity fields $\delta(\mathbf{x}) = (\rho(\mathbf{x}) - \bar{\rho}) / \bar{\rho}$ extracted from the fiducial cosmology \texttt{Quijote} $N$-body simulation suite \citep{2020ApJS..250....2V}. These simulations track the evolution of $512^3$ cold dark matter particles in a periodic cubic volume of $V = L^3 = (1 \ \mathrm{Gpc} / h)^3$, running from $z = 127$ to $z = 0$. They employ the \texttt{Gadget-III} TreePM code \citep{Springel:2005mi}, varying random phases of the initial conditions and keeping the cosmological parameters fixed to the following fiducial Planck values: $\Omega_{\mathrm{m}}=0.3175, \Omega_{\mathrm{b}}=0.049, h=0.6711, n_s=0.9624, \sigma_8=0.834, M_\nu=0.0 \ \mathrm{eV}$, and $w=-1$ \citep{2020A&A...641A...6P}. The fields are produced by meshing the particles on a grid with resolution of $N=N_{\rm g}^3 = 128^3$ via the piecewise cubic spline mass assignment scheme. An example simulation can be seen in Fig.~\ref{fig:samples}, which is our ground truth fiducial simulation that we analyse and which is not contained in the training/validation set. Each such simulation takes $\sim 10^3$ CPU hours to complete.

\subsection{Statistical problem} \label{subsec:statistical_problem}

Our observational data, $\bfx$, is given by the final ($z=0$) dark matter density field, and our objective is to infer the distribution of initial ($z=127$) dark matter density fields, $\bfz$, which are consistent with this observation $\bfx$. For the results presented herein, we do not add any noise to the observational data $\bfx$, but we verified that small amounts of noise do not change the outcomes significantly. Both $\bfx$ and $\bfz$ contain $128^3$ values on a regular grid, which in total amounts to $N \approx 2 \cdot 10^6$ parameters whose distribution we aim to infer. We formulate this problem in the Bayesian framework:
\begin{equation}
\label{eq:bayes}
    p(\bfz|\bfx)=\frac{p(\bfx|\bfz)}{p(\bfx)}p(\bfz),
\end{equation}
where $p(\bfz)$ is the prior distribution, $p(\bfx|\bfz)$ is the likelihood, $p(\bfx)$ is the evidence, and our goal is to generate samples from the posterior distribution $p(\bfz|\bfx)$. 

According to the standard cosmological model, the complex, highly non-Gaussian observed present-day LSS formed in the process of gravitational collapse over billions of years of cosmic evolution from tiny primordial fluctuations around a highly homogeneous and isotropic state. Observational data from the Cosmic Microwave Background (\citealt{2020A&A...641A...6P}) show that these fluctuations at high redshifts are well described by a Gaussian random field. Therefore, we can write down the prior distribution $p(\bfz)$ for the initial $z=127$ field in the form:
\begin{equation}
\label{eq:prior}
    p(\bfz) = \cfrac{\exp{-\frac12 \bfz^T \bm Q^P \bfz}}{\sqrt{(2\pi)^N (\det (\bm Q^P)^{-1})}} \, ,
\end{equation}
where the prior precision matrix $\bm Q^P$ is diagonal in Fourier space:
\begin{equation}
\label{eq:Q_prior}
    \bm Q^P = \bm U^{\dagger} \bm D^P \bm U \, .
\end{equation}
Here, $\bm U$ and $\bm U^{\dagger}$ denote the discrete Fourier transform and its inverse, respectively, and the diagonal matrix is given by $\bm D^P = 1/P(k)$, with $k \equiv |\bm k|$  and $P(k)$ being the linear matter power spectrum at $z = 127$, which we compute using the \texttt{CLASS} code \citep{Lesgourgues:2011re, Blas:2011rf}. In practice, our implementation uses the discrete Hartley transform $\bm H$ rather than the discrete Fourier transform. The Hartley transform is closely related to the Fourier transform: it is also unitary and diagonalises the precision matrix, but has the additional advantage that real values remain real, avoiding the need to deal with complex numbers at the code implementation level while not changing any of the mathematical arguments presented herein (see App.~\ref{appendix_hartley}). As our underlying framework works for both the standard Fourier transform and the Hartley transform, however, we will often omit this subtle distinction when describing our method in what follows.

As density perturbations grow over cosmic time, the nonlinear nature of gravity causes the evolution of different Fourier modes to couple, and the density field gradually becomes non-Gaussian. Thus, a key choice for any Bayesian cosmological reconstruction framework lies in the level of complexity used for the description of the likelihood (see e.g.\ the elaborate `multi-scale likelihood model' devised by \citealt{Doeser:2023yzv}).

Usually, in the context of SBI, the form of the likelihood or posterior is not fixed, but rather parametrised by a neural network and hence learned from the data in the process of training. In order to make inference for the high-dimensional IC reconstruction problem tractable, however, we adopt a simpler approach. In particular, we fix the parametric form of the likelihood to be Gaussian in the inferred initial conditions $\bfz$ with an $\bfx$-dependent mean $\hat\bfz_\bftheta(\bfx)$ and a precision matrix $\bm Q_\bftheta^L$, which we set to be Fourier-diagonal because both the statistical properties of the initial field and the physics of structure formation are translationally invariant. Note that our simulation box has periodic boundary conditions.

Our likelihood, up to a normalisation factor, is therefore given by
\begin{equation}
\label{eq:likelihood}
    p(\bfx|\bfz) \propto \exp{-\frac12 (\bfz - \hat\bfz_\bftheta(\bfx))^T \bm Q_\bftheta^L (\bfz- \hat\bfz_\bftheta(\bfx))},
\end{equation}
where $\bm Q_\bftheta^L = \bm U^{\dagger} \bm D_{\bftheta}^L \bm U$ with diagonal $\bm D_{\bftheta}^L$, $\hat{\bfz}_\bftheta(\bfx)$ is the $\bfx$-dependent maximum likelihood estimator (MLE) and the subscript $\bftheta$ denotes quantities that depend on the set of trainable parameters of the model  and are learned during training. 

Perhaps surprisingly, this simple approach is sufficient to obtain a tight and statistically consistent posterior distribution in the chosen range of scales, 
as we will demonstrate below. We expect this approximation to work well on very large scales, where the evolution of modes is simple and linear, and on very small scales, where gravitational collapse erases all the ability to recover detailed physical information. However, this ansatz is only an approximation
on the intermediate scales where the modes become highly coupled but the information recovery is still possible. In particular, by accounting for the cross-correlations between the Fourier modes present in the observed data only at the level of the mean estimator and neglecting it at the level of the precision matrix $\bm Q_\bftheta^L$, we leave some of the information unused, resulting in conservative, rather than overconfident, posteriors.

\begin{figure*}
    \includegraphics[width=18cm]{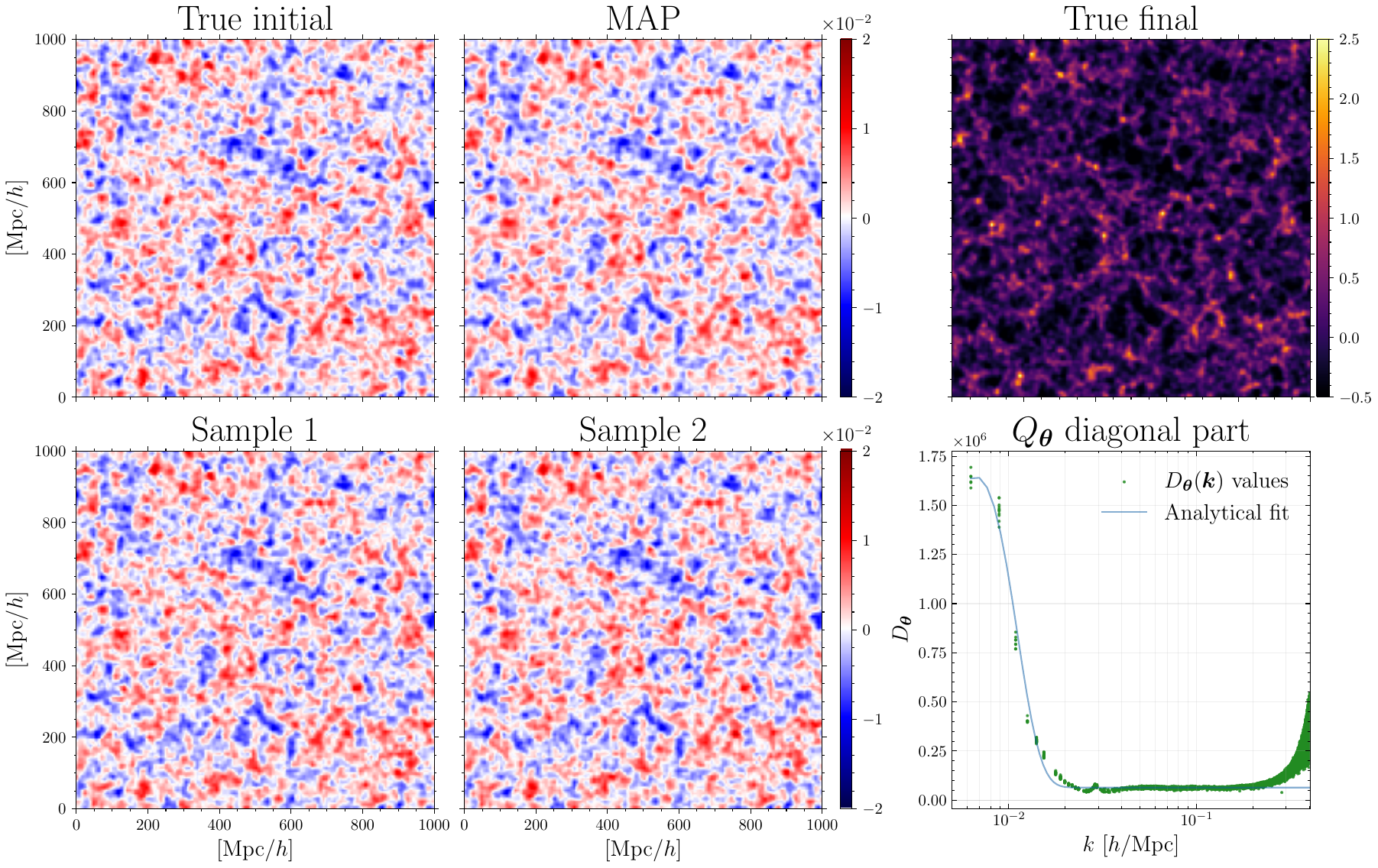}
    \caption{Slices of the results of our Gaussian neural posterior estimation for 3D reconstruction of cosmological ICs. \textit{Top left}: ground truth initial overdensity fields $\bfz_{\mathrm{truth}}$. \textit{Top centre}: MAP estimate of the initial condition given the observed final conditions $\bfx_{\mathrm{obs}}$ computed with the estimator $\mathbf{\hat{\boldmu}}_\bftheta(\bfx_{\mathrm{obs}})$. \textit{Bottom left and centre}: two examples of the generated posterior ICs samples. \textit{Top right}: given late-time density field, i.e.\ the observation $\bfx_{\mathrm{obs}}$. All the shown slices are averaged over the depth of $100 \ \text{Mpc}/h$ in the third axis direction. \textit{Bottom right}: the posterior precision matrix is modelled to be diagonal in Fourier space: $\bm Q_\bftheta = \bm Q^P + \bm Q_\bftheta^L = \bm U^{\dagger} \bm D_{\bftheta} \bm U$. Here we show the diagonal component $\bm D_{\bftheta} = \bm D^P + \bm D_{\bftheta}^L$ trained values as a function of the wavenumber $k$ ({\it green dots}), as well as a simple analytical Gaussian fit of the $\bm D_{\bftheta}(k)$ dependence accurate up to $k \simeq 0.2 \, h/\text{Mpc}$ ({\it blue line}). Knowing the form of the $\bm D_{\bftheta}(k)$ function allows one to turn any ICs point estimator into a fast sampler, as we discuss in Section~\ref{subsec:fit}. The precision is indicative of the confidence in the reconstructed ICs at a given scale and therefore peaks at large scales, which are easier to infer.}
    \label{fig:samples}
\end{figure*}

Plugging equations (\ref{eq:prior}) and (\ref{eq:likelihood}) into Bayes' theorem (\ref{eq:bayes}) gives rise to the following posterior distribution:
\begin{equation}
\label{eq:posterior}
    p(\bfz|\bfx) = \cfrac{\exp{-\frac12 (\bfz - \mathbf{\hat{\boldmu}}_\bftheta(\bfx))^T (\bm Q^P + \bm Q_\bftheta^L)(\bfz - \mathbf{\hat{\boldmu}}_\bftheta(\bfx))}}{\sqrt{(2\pi)^N \det(\bm Q^P + \bm Q_\bftheta^L)^{-1}}} \, ,
\end{equation}
where $\mathbf{\hat{\boldmu}}_\bftheta(\bfx)$ is the MAP estimator related to the MLE by $\mathbf{\hat{\boldmu}}_\bftheta(\bfx) = (\bm Q^P + \bm Q_\bftheta^L)^{-1} \bm Q_\bftheta^L \, \hat\bfz_\bftheta(\bfx)$. Formula (\ref{eq:posterior}) is our main ansatz, and our goal is to train the $\bfx$-dependent MAP estimator $\mathbf{\hat{\boldmu}}_\bftheta(\bfx)$ and the $\bfx$-independent likelihood precision matrix diagonal values $\bm D_{\bftheta}^L$. Note that, as can be seen from (\ref{eq:posterior}), the Gaussian form of the prior and the likelihood results in a Gaussian posterior, whose precision matrix is simply the sum of the prior and the likelihood precision matrices: $\bm Q_\bftheta = \bm Q^P + \bm Q_\bftheta^L$. This allows one not to explicitly prescribe any prior information and directly train the posterior quantities; we found that training the posterior diagonal values $\bm D_{\bftheta}$ in this way leads to essentially the same results.

Before training, we rescale the $z=127$ fields by the inverse of the linear growth factor $\mathcal{D}(z=127) \simeq 9.9 \times 10^{-3}$, such that the power spectrum of the initial and final fields agree at large scales.\footnote{In order to match the \texttt{Quijote} conventions, we take the approximate analytical expression for $\mathcal{D}(z)$ from \citealt{Carroll:1991mt}.} This transformation effectively makes the network learn the residuals from the linear evolution, which we found to improve the training significantly. After training, it is straightforward to rescale the MAP estimator $\mathbf{\hat{\boldmu}}_\bftheta(\bfx)$ and the precision matrix $\bm Q_\bftheta$ back to the original field normalisation.

We implement the $\bfx$-dependent MAP estimator $\mathbf{\hat{\boldmu}}_\bftheta(\bfx)$ using a 3D U-Net type neural network architecture \citep{ronneberger2015u}, which we take from the \texttt{map2map} package \citep{Jamieson:2022lqc}, simplifying it by using 16 instead of 64 hidden channels. A U-Net-style neural network performs convolutions across multiple resolution scales, forming a U-shaped structure with a sequence of downsampling layers followed by a matching number of upsampling layers. Such a configuration allows effective capturing of detailed spatial features, and the presence of skip connections enables processing information at various scales, making it especially well-suited for analysing data like the cosmological LSS. To account for the periodicity of the simulation box, periodic padding is used for the convolutional layers.

To guide the network in accurately reconstructing the ICs field across all scales, we model $\mathbf{\hat{\boldmu}}_\bftheta(\bfx)$ as a sum of two terms, effectively modifying the last activation layer of the neural net in such a way that the U-Net is tasked with reconstructing only the small scales, while the evolution of large scales is described by a simple multiplicative scaling factor $\bfalpha_\bftheta(\bfk)$:
\begin{equation}
\label{eq:MAP}
    \mathbf{\hat{\boldmu}}_\bftheta(\bfx) = \bm U^{\dagger} \left\{ \bfalpha_\bftheta(\bfk) \odot \left[\bm U\{\bfx\} + \sigma_{> k_\Lambda}(\bm U\{\operatorname{U-Net}_\bftheta(\bfx)\})\right] \right\}.
\end{equation}
Here, $\sigma_{> k_\Lambda}$ is a sigmoidal high-pass filter centred at the cut-off $k_{\Lambda} = 0.03 \, h/\text{Mpc}$ \footnote{This choice of cut-off is motivated from the scale at which the cross-correlation between initial and final conditions starts to noticeably deviate from 1 and so non-linear effects start to kick in (see the orange $C(k)$ line in Fig. \ref{fig:sum}).}, $\bfalpha_\bftheta(\bfk)$ corresponds to a trainable scaling factor applied to each Fourier mode, and $\odot$ is the Hadamard (elementwise) product. We found that including this last custom sigmoidal activation function leads to improved results, as it allows the U-Net to limit its attention to non-linear scales (without this layer, the inferred fields turned out to have biased power spectra at large scales). While the scaling factors $\bfalpha_\bftheta(\bfk)$ are not essential to our method and can also be omitted (in particular when applying our method to higher-resolution boxes, where the additional storage of $N$ numbers might be undesirable), we observed that incorporating $\bfalpha_\bftheta(\bfk)$ leads to a slight improvement in performance, without any major effects on the training speed. Note that our two-scale ansatz for the MAP in Eq.\ (\ref{eq:MAP}) is conceptually similar to the Hybrid SBI approach of \citet{Modi:2023drt}, which combines large-scale information from perturbation theory with small-scale information from simulations.

The second trainable part of our model, the diagonal entries of the likelihood precision matrix $\bm D_{\bftheta}^L$, are expected to depend only on the absolute value $k$ of each Fourier mode $\bm k$ due to the physical rotational symmetry, i.e.\  $\bm D_{\bftheta}^L = \bm D_{\bftheta}^L (k)$.
We parametrised $\bm D_{\bftheta}^L$ and performed the training in two different ways: (1) leaving the values $\bm D_{\bftheta}^L (\bm k)$ as free parameters, or (2) enforcing rotational symmetry from the start by implementing $\bm D_{\bftheta}^L$ as a three-layer dense neural network with 128 hidden channels which takes $k$ as an input and outputs the $\bm D_{\bftheta}^L$ value. In both cases, in order to ensure positivity, we actually parametrise the square root of $\bm D_{\bftheta}^L$ and then square it at inference. We found the two methods to perform similarly overall, although we observed slightly better precision and more stable training for the `free' $\bm Q_\bftheta^L$ matrix (suspectedly, because the free matrix can compensate for rotationally asymmetric behaviour of the MAP estimator, whose convolutional kernels are not constrained to be rotationally symmetric). In this case, the rotational symmetry manifests itself in the small amount of scatter of $\bm D_{\bftheta}^L$ values for a given $k$ (see the bottom right corner of Fig. \ref{fig:samples}.) We provide an analytical fit of the $\bm D_{\bftheta}^L(k)$ dependence for the `free' $\bm Q_\bftheta^L$ matrix and draw its practical applications in Section~\ref{subsec:fit}.

\subsection{Training strategy} \label{subsec:training}

Minimising the negative log posterior probability of observing the training data under the ansatz (\ref{eq:posterior}) yields the following training objective:
\begin{multline}
\label{eq:loss}
    \mathcal{L}  = - \sum_{i=1}^n \log p(\bfz_i|\bfx_i) = \frac12 \sum_{i=1}^n \left((\bfz_i - \mathbf{\hat{\boldmu}}_\bftheta(\bfx_i))^T \bm Q_\bftheta \,(\bfz_i - \mathbf{\hat{\boldmu}}_\bftheta(\bfx_i))\right) \\
   - \frac{n}{2} \tr \log \bm Q_\bftheta - \frac{N n}{2} \log 2 \pi, \hspace{2.3cm}
\end{multline}
where $N=N_{\rm g}^3$, the index $i$ runs over the training set, $n$ is the number of samples in the training set (or in a batch), $\bm Q_\bftheta = \bm Q^P + \bm Q_\bftheta^L$, and we used the $\log \det \bm Q_\bftheta = \tr \log \bm Q_\bftheta$ identity. Note the similarity of our method to mean-field approximate models encountered in the contexts of statistical physics (where many-body interactions are approximated by an average field, e.g.\ \citealt[Chap.\ 4]{yeomans1992statistical}) and variational inference (where the approximate distribution is factorised into independent marginals of the latent variables for computational tractability, e.g.\ \citealt{Blei_2017}).Moreover, our approach aligns with the SBI framework in \citet{jeffrey2020solving}, where a hierarchy of networks estimates increasing moments of the posterior distribution; hence, focusing on the first two moments, i.e., mean and variance, relates to our Gaussian NPE method (note, however, the difference in the training objective and the learning strategy).

We randomly split the training dataset, allocating 80\% for training and 20\% for validation. We train the model with a batch size of 8 on a single 40 GB $\texttt{Nvidia A100}$ GPU with an initial learning rate of $10^{-2}$, and we decrease the learning rate by a factor of 10 if the validation loss plateaus for one epoch. We train the model for 30 epochs until we see no further decrease in the validation loss, with the loss plateauing for the last $\sim 10$ epochs, which under these settings takes approximately 1.5 hours to complete. For inference, we load the weights with the minimum validation loss. To implement the neural network and training, we use the \texttt{PyTorch Lightning}-based \citep{PyTorch_Lightning_2019} code \texttt{swyft} \citep{Miller:2021hys, Miller:2022shs}. We use \texttt{Zarr} \citep{zarr} to effectively store the training data. Our code will be made publicly available here: \href{https://github.com/oleg-savchenko/gaussian-npe}{\texttt{https://github.com/oleg-savchenko/gaussian-npe}}.

\begin{figure}
    \includegraphics[width=\columnwidth]{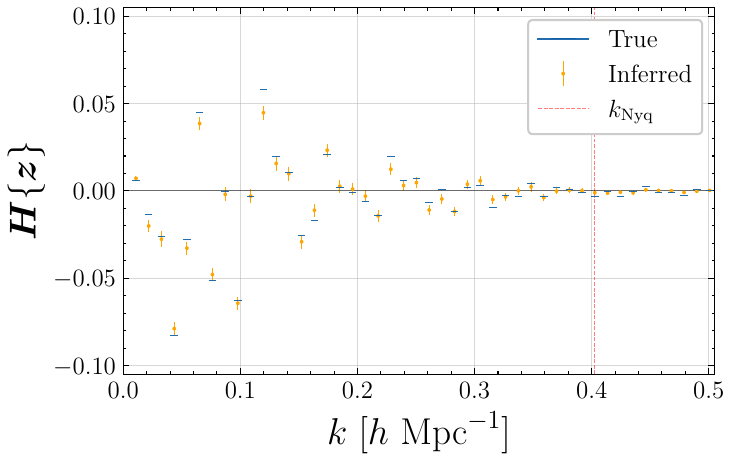}
    \caption{The trainable parts of the algorithm are the MAP estimator $\mathbf{\hat{\boldmu}}_\bftheta$ and the likelihood precision matrix $\bm Q_\bftheta^L$. Orange vertical lines depict the mean of the posterior predictions for some randomly selected individual Hartley modes, equal to $\bm H \{\mathbf{\hat{\boldmu}}_\bftheta(\bfx_{\mathrm{obs}})\}(\bfk)$ (with $\bm H$ denoting the Hartley transform, see App. \ref{appendix_hartley}), together with their $1\sigma$ errors, equal to $(\bm D^P + \bm D_\bftheta^L)^{-\frac{1}{2}}$. Horizontal blue lines show the true value of a given mode.}
    \label{fig:modes_uncertainties}
\end{figure}

\section{Results} \label{sec:results}
In this section, we present the outcome of our Gaussian NPE approach. First, we showcase the generated posterior samples, highlighting their visual consistency with the true initial conditions (Section~\ref{subsec:samples}). We then compare a range of summary statistics between the generated samples and ground truth (Section~\ref{subsec:stat}). Next, we verify the coverage properties of our posterior on different scales via a Bayesian consistency test, ensuring that the inferred uncertainties are well-calibrated across various $k$-bins (Section~\ref{subsec:coverage}). Finally, we examine the structure of the trained likelihood precision matrix in detail, demonstrating how a simple functional form can enable rapid sampling from high-dimensional posteriors (Section~\ref{subsec:fit}).

\subsection{Samples} \label{subsec:samples}
Once training has converged, the method allows extremely fast sampling from the posterior since it merely requires drawing samples from a high-dimensional multi-variate Gaussian with a diagonal covariance matrix and transforming them to real space via a fast Fourier transform. Examples of two generated IC samples for a fiducial \texttt{Quijote} simulation $\left\{\bfz_{\mathrm{truth}}, \bfx_{\mathrm{obs}}\right\}$ not contained in the training/validation set is shown in Fig.~\ref{fig:samples}. Visual inspection indicates a high level of resemblance to the ground truth. Given this input final field $\bfx_{\mathrm{obs}}$, it takes $\lesssim 1 \, \mathrm{s}$ of computational time to obtain the IC MAP estimate $\mathbf{\hat{\boldmu}}_\bftheta(\bfx_{\mathrm{obs}})$ via (\ref{eq:MAP}). The trained posterior precision matrix $\bm Q_\bftheta = \bm Q^P + \bm Q_\bftheta^L$ then quantifies the mode-by-mode reconstruction precision.

Posterior samples are then drawn as $\bfz = \mathbf{\hat{\boldmu}}_\bftheta(\bfx_{\mathrm{obs}}) + \bm H\{\sigma(\bm k) \odot \boldsymbol{\varepsilon}\}$,  
where $\boldsymbol{\varepsilon}$ is a random white noise field sample and $\sigma(\bm k) = (\bm D^P + \bm D_\bftheta^L)^{-\frac{1}{2}}$. Predicted means and uncertainties for some individual modes can be seen in Fig. \ref{fig:modes_uncertainties}. Generating 1000 samples in a batched manner in this way takes $\lesssim 3 \, \mathrm{s}$ on a single GPU.

\subsection{Summary statistics comparison} \label{subsec:stat}
\begin{figure}
    \includegraphics[width=\columnwidth]{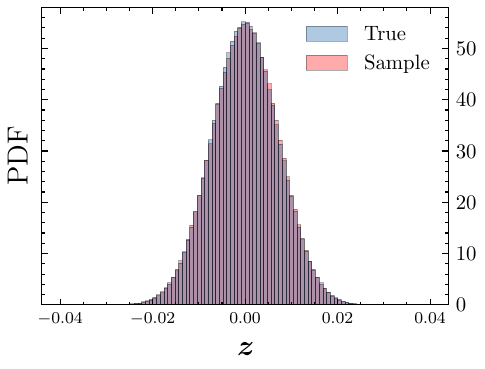}
    \caption{Comparison of the density field histograms (1-point functions) for the true field $\bfz_{\mathrm{truth}}$ and one of the generated samples.}
    \label{fig:1-pt}
\end{figure}

\begin{figure}
    \includegraphics[width=\columnwidth]{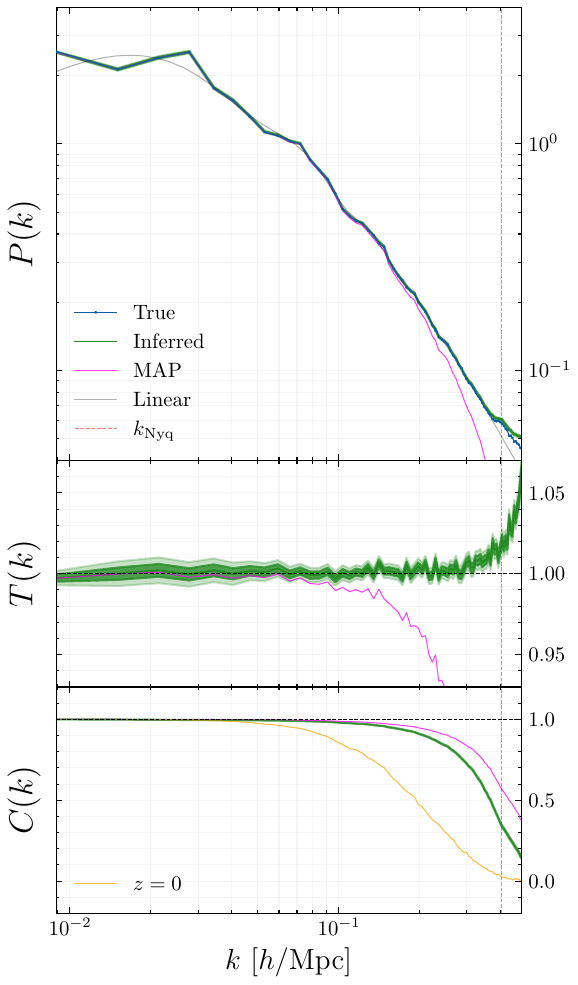}
    \caption{Comparison of the summary statistics at $z=127$ between the ground truth simulation and the generated samples together with the MAP estimate. \textit{Top}: power spectra $P(k)$. \textit{Middle}: transfer function $T(k)$. \textit{Bottom}: cross-power spectra $C(k)$. For summaries of the generated samples, solid lines represent the mean, while shaded regions correspond to $1\sigma$ and $2\sigma$ uncertainties computed from a 1000 generated samples using the \texttt{Pylians3} package \citep{2018ascl.soft11008V}. The plots demonstrate that generated samples reproduce the true field with a high level of precision.}
    \label{fig:sum}
\end{figure}

To assess the physical consistency of the produced dark matter density field ICs samples and to quantify the accuracy of the obtained results, we compute their 1-point function (see Fig. \ref{fig:1-pt}), as well as their power spectrum, transfer function {$T(k)$}, and cross-correlation {$C(k)$} relative to the ground truth field $\bfz_{\mathrm{truth}}$ (see Fig. \ref{fig:sum} and App. \ref{appendix_summary} for definitions). The generated samples show a high level of similarity with the ground truth field: we observe an agreement in the power spectrum at the level of $\lesssim 1-2\%$ all the way up to the Nyquist scale $k_{\mathrm{Nyq}} \simeq 0.4 \, h/\mathrm{Mpc}$ and a high ($\gtrsim 70\%$) cross-correlation up to $k \simeq 0.3 \, h/\mathrm{Mpc}$. 

Note that, unlike for point estimators studied in most other works, our generated fields do not lose significant power at small scales and thus more accurately capture the ICs field statistical properties. The high level of cross-correlation compared to the $C(k)$ computed for the $z=0$ field  further demonstrates that the neural network is able to capture and disentangle the highly non-linear mode coupling of the final field. Moreover, we verified that our approach also works when we add Gaussian noise to the final observed field: for $\sigma=0.1$ noise we observe the same level of power spectrum agreement, and the cross-correlation drops to 20\% at $k_{\mathrm{Nyq}}$ (as compared to 35\% without added noise).

\subsection{Scale-dependent consistency test} \label{subsec:coverage}
\begin{figure*}
    \includegraphics[width=1.8\columnwidth]{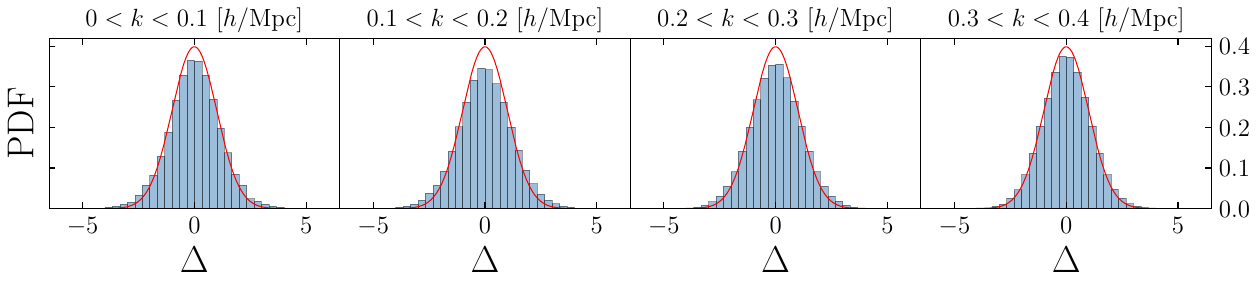}
    \caption{Bayesian consistency check for different wavenumber bins. If samples follow the correct posterior distribution, the resulting histograms should follow the $\mathcal{N}(0, 1)$ univariate Gaussian distribution.}
    \label{fig:coverage}
\end{figure*}

In order to verify that the generated samples indeed exhibit proper coverage with respect to the true posterior, we perform a Bayesian consistency check and study the distribution not only for all voxels taken in aggregate, but also for subsets of modes at different scales (see Fig.~\ref{fig:coverage}).

Specifically, we assess whether the inferred IC samples are statistically consistent with the true posterior distribution within distinct $k$-bins. 
We begin by defining the Fourier transform of the true ICs as $\tilde{\bfz}_\mathrm{truth} := \bm U \{\bfz_\mathrm{truth}\}$  and the mean of the Fourier-transformed posterior samples as $\bar{\tilde{\bfz}}_\mathrm{samples} \equiv \mathbb{E}[\tilde{\bfz}_\mathrm{samples}]$. We then normalise the difference between $\tilde{\bfz}_\mathrm{truth}$ and $\bar{\tilde{\bfz}}_\mathrm{samples}$ by the standard deviation of the Fourier transforms of the posterior samples, $\tilde{\sigma}_\mathrm{samples} \equiv \sqrt{\mathbb{E}[(\tilde{\bfz}_\mathrm{samples})^2] - \left(\mathbb{E}[\tilde{\bfz}_\mathrm{samples}]\right)^2}$, where $\bfz_\mathrm{samples}\sim p(\bfz| \bfx)$ {as modelled in Eq.\ (\ref{eq:posterior})}. Formally, we define the quantity 
\begin{equation}
\label{eq:coverage}
    \Delta(\bm k) \equiv \cfrac{
    \tilde{\bfz}_\mathrm{truth} (\bm k) - \bar{\tilde{\bfz}}_\mathrm{samples} (\bm k)}{\tilde{\sigma}_\mathrm{samples} (\bm k)} \;.
\end{equation}
Because our likelihood precision matrix is nearly rotationally symmetric (as expected from isotropy), we group modes within spherical shells in $k$-space and examine the consistency of $\Delta(\bm k)$ within each bin. Fig.~\ref{fig:coverage} shows that these normalised differences follow a univariate normal distribution with mean zero and unit variance across the various $k$-ranges, demonstrating that our posterior samples capture the first two moments of the true distribution well. This constitutes a more stringent test than that of \citet[Fig.~5]{Legin:2023jxc}, which evaluated coverage collectively across all modes, whereas here we confirm statistical consistency separately for \textit{each individual $k$-bin}.

\subsection{Study of the likelihood precision matrix structure} \label{subsec:fit}

As we mentioned before, physical isotropy manifests itself in the fact that the trained likelihood precision matrix $\bm Q_{\bftheta}^L$ is approximately rotation-invariant and its diagonal values $\bm D_{\bftheta}^L$ approximately only depend on the magnitude of the wavevector $k$ (see bottom right of Fig.~\ref{fig:samples}). We verified that this makes it possible to impose this symmetry from the start by parametrising $\bm D_{\bftheta}^L$ as a function of $k$ via a dense neural network, with only slightly less precise results. In addition, we found that for all the $k$ values up to $k_{\mathrm{Nyq}}$ we have $|\bm D_{\bftheta}^L| \gg |\bm D^P|$, so in this sense the posterior is dominated by information coming from the likelihood. A tiny bump is visible at the cutoff scale $k_\Lambda = 0.03 \ h / \mathrm{Mpc}$ (see Eq.\ \ref{eq:MAP}); however, this does not noticeably affect the quality of the posterior samples.

Here, we perform a simple analytical fit of the posterior precision matrix diagonal values $\bm D_{\bftheta}$ for the case of the `free' (unconstrained) $\bm Q_\bftheta^L$ matrix as a function of the wavenumber $k$ up to $k = 0.2 \, h/\text{Mpc}$ in form of a Gaussian with 4 fitting parameters:
\begin{equation}
\label{eq:fit}
    D_{\bftheta, \, ii}(k) = A \cdot \exp\left(-\frac{(k - \mu)^2}{2 \sigma^2}\right) + C \, ,
\end{equation}
which can describe $\bm D_{\bftheta}$ quite accurately up to $k \simeq 0.2 \, h/\text{Mpc}$, but doesn't capture the growing behavior of $\bm D_{\bftheta}$ that occurs at smaller scales close to $k_{\mathrm{Ny}}$. We expect this growing behaviour to indicate the presence of numerical (hence, non-physical) artefacts in the field, which allow the method to be overly confident in its predictions on these scales. Therefore, the exact shape close to $k_{\mathrm{Nyq}}$ will depend on the numerical details of the specific $N$-body simulation and other aspects, such as the mass assignment scheme used to obtain the fields. More generally, it would be interesting to study the sensitivity of these fitted parameters with respect to cosmology and the numerical details of the simulation and the density computation; we leave this exploration for future work.

We find the following values for the parameters of the fit (\ref{eq:fit}): $A = 1.58 \times 10^6, C = 6.43 \times 10^4, \mu = 6.67 \times 10^{-3} \, h/{\rm{Mpc}}, \sigma = 3.76 \times 10^{-3} \, h/{\rm{Mpc}}$ (see the fit curve in the bottom right panel of Fig. (\ref{fig:samples})). So, given a trained MAP estimator, our proposed algorithm for transforming it into a fast ICs sampler is the following:
\begin{enumerate}[label = (\arabic*), leftmargin=2.5em, labelindent=1em]
    \item Train a MAP point estimator, using e.g. a mean-square error (MSE) loss function.
    \item Generate a Gaussian random field sample in Fourier space with the analytic fit precision matrix \eqref{eq:fit}, and transform it back to real space.
    \item Apply the trained MAP estimator to a given observed field $\bfx_{\mathrm{obs}}$ and then add the generated Gaussian random field to it.
\end{enumerate} 
Samples generated with this analytically fitted $\bm Q_{\bftheta}$ matrix exhibit the same level of agreement in cross- and power-spectrum up to $k \simeq 0.2 \, h/\text{Mpc}$ compared to the fully trained ansatz (\ref{eq:posterior}), as we demonstrate in App. \ref{appendix_mse}.

Even though different MAP estimators might be trained with different loss functions, any reasonable loss function which is minimised by the MAP estimator should eventually converge and give rise to the same resulting mapping. Notably, imposing a diagonal precision in our Gaussian ansatz ($\ref{eq:posterior}$) reduces the number of free parameters in this matrix from $N \times N \sim 10^{12}$ to $N \sim 10^6$. Additionally assuming rotational invariance further shrinks this number to just $\mathrm{\sim} \, 10^2$ values on a grid in $k$-space, and by finding a simple analytic fit (\ref{eq:fit}), we eventually reduced this just to a few parameters.

\section{Discussion \& Conclusions} \label{sec:conclusion}

In this work, we developed and implemented a simple yet effective method to obtain samples from the posterior of initial conditions of the Universe's dark matter density field conditioned on the observation of the final density field. Our main assumption --- that this posterior can be modelled as a Gaussian distribution in the inferred ICs $\bfz$ with a diagonal precision matrix in Fourier space --- enables a large reduction in model complexity. The trainable parameters consist of the observation-dependent mean estimator, which we model using a U-Net neural network with a physics-informed custom final activation layer, and the diagonal values of the posterior precision matrix $\bm D_{\bftheta} = \bm D^P + \bm D_{\bftheta}^L$, which give the precision of reconstruction of a given Fourier mode. After training, which takes $\mathrm{\sim}1.5 \, \mathrm{h}$ on a single GPU, our method not only yields point estimates (as is common with many neural-network–based approaches), but also provides a simple recipe for generating posterior samples in an extremely fast way by drawing them from a high-dimensional multivariate Gaussian ($\mathrm{\lesssim} \, 3 \, \mathrm{s}$ sampling time on a GPU for 1000 samples). In addition, our SBI method does not require the forward model to be differentiable (unlike HMC-based approaches), and has the property of being fully {\it amortised}, meaning that the trained model can be applied to any given observation and produce samples for it right away.

By calculating various summary statistics of the generated samples, we verify that they have the correct statistical properties and are consistent with the ground truth up to a high precision (the power spectrum agrees within $\mathrm{\lesssim} \, 1\mathrm{-}2\%$ up to the Nyquist scale $k_{\mathrm{Nyq}} \simeq 0.4 h/\mathrm{Mpc}$, with a high cross-correlation ($\mathrm{\gtrsim} \, 70\%$) persisting up to $k \simeq 0.3 \, h/\mathrm{Mpc}$). To further ensure the proper coverage of the posterior across different scales, we conduct a Bayesian consistency test for modes within different $k$-bins. Moreover, we determine a simple analytic fit of the found precision $\bm D_{\bftheta} (k)$ dependence, which allows turning any trained point estimator into a fast posterior sampler.
 
Although the Gaussian Fourier-diagonal approximation for the posterior proved sufficient in the selected range of scales $k \lesssim \, 0.4 \, h/\rm{Mpc}$, applying it to smaller scales, or including observational effects which break translational invariance (e.g. redshift-space distortions, space-dependent noise), might necessitate a more sophisticated precision matrix, or potentially moving beyond the Gaussian approximation entirely. As an example of an alternative basis for the precision matrix, we experimented with the convolutional matrix $\bm Q_\bftheta^L = \mathcal{C}_{\bftheta}^{\dagger} \bm D_{\bftheta}^L \mathcal{C}_{\bftheta}$, where $\bm D_{\bftheta}^L$ is diagonal and $\mathcal{C}_{\bftheta}$ denotes a convolution operation of a certain kernel size, achieving altogether similar reconstruction results. Instead of convolutions, one can also consider wavelets, the scattering transform, etc. In those cases, the prior and the likelihood precision matrices are diagonal in different bases, so sampling is no longer as straightforward and fast, and one needs to resort to techniques such as data augmentation approaches for sampling (see Appendix~B of \citealt{List:2023jwo}). In addition, a more realistic precision matrix is expected to depend on the observation, $Q_\bftheta^L (\bfx)$, to reflect the fact that the reconstruction variance correlates with high-density regions.

It would also be interesting to simplify the structure of our MAP estimator by optimising the U-Net architecture, removing the dependency on the choice of cut-off scale $k_{\Lambda}$. These improvements should enhance the generality of our framework without compromising speed and accuracy, making it suitable for a broader range of field-level reconstruction problems. The range of scales on which the ICs can be reconstructed with our method can also be extended to smaller scales simply by using simulations meshed on a higher resolution box (e.g. $256^3$ or more), which, however, would require more computational resources to train the model (see the comparison e.g. in \citealt{Floss:2023ylq}). That said, it would be interesting to study, by means of higher-resolution simulations, how the reconstruction of cosmic ICs is {\it fundamentally} limited by the non-linear and increasingly chaotic dynamics when the Nyquist frequency is gradually increased. While \citet{Legin:2023jxc} suggest that their score-based model nearly saturates the entire information content regarding the ICs in the late-time field (note that their reconstruction quality is very similar to ours in terms of (cross-)power spectrum), this hypothesis is called into question by the results in \citet[Fig.\ 3]{Floss:2023ylq}, who show that the reconstruction quality can be further improved by increasing the resolution from $128^3$ to $256^3$, also for the  $(1 \ \mathrm{Gpc} / h)^3$ \texttt{Quijote} box. We defer a detailed study in this direction to future work.

Our approach opens up numerous promising opportunities for future investigation. While we trained the model with fixed cosmological parameters, it will be very interesting to test it for varying cosmology and infer the parameters alongside the cosmic ICs. For this task, a promising enhancement of our method would involve leveraging the sequential aspect of SBI techniques~\cite[e.g.,][]{2018arXiv180507226P, Miller:2021hys,  2022arXiv221004815D}. Moreover, we have tested our method under an idealised scenario where the observation is represented by the dark matter density field. In practice, what we observe are galaxies, which are baryonic biased tracers of the dark matter field, and additionally the observations are subject to many systematic effects (e.g. redshift space distortions, lightcone effects, survey masks, etc). In future work, we plan to incorporate these observational and physical effects by means of a hierarchical simulator, so that our method could be ultimately applied to real data.

\section*{Acknowledgements}

OS, GFA, NAM and CW acknowledge support from the European Research Council (ERC) under the European Union’s Horizon 2020 research and innovation programme (Grant agreement No. 864035 - Undark). The main analysis for this work was carried out on the Snellius Computing Cluster at SURFsara. We are grateful to Uddipta Bhardwaj, Mathis Gerdes, and Thomas Flöss for useful discussions. FL thanks Oliver Hahn for many insightful conversations.

We acknowledge the use of the following software: \texttt{NumPy} \citep{Harris:2020xlr}, \texttt{Matplotlib} \citep{Hunter:2007ouj}, \texttt{Pylians3} \citep{2018ascl.soft11008V}, \texttt{PyTorch} \citep{Paszke:2019xhz}, \texttt{PyTorch Lightning} \citep{PyTorch_Lightning_2019}, and \texttt{Jupyter} \citep{jupyter}.





\bibliographystyle{mnras}
\bibliography{bibliography} 




\appendix

\section{Summary statistics}
\label{appendix_summary}

Given two (dimensionless) 3D matter overdensity fields $\delta_{a, b}(\mathbf{x})=\rho_{a, b}(\mathbf{x})/\bar{\rho}_{a, b}-1$, one can define their Fourier-space 2-point correlation function by
\begin{equation}
\label{eq:Pk_def}
    \langle \tilde{\delta}_a (\bfk) \tilde{\delta}_b^{*} (\bfk') \rangle = (2\pi)^3 \, P_{a b}(k) \, \delta^{(3)}(\bfk - \bfk') \, .
\end{equation}
Here, $\tilde{\delta}_{a, b}(\bfk) = \bm U \{\delta_{a, b}(\mathbf{x})\}$ is the fields' Fourier transform, the brackets $\langle ...\rangle$ denote an average over many realisations, and the 3D Dirac delta function arises from translational invariance, while the fact that $P_{a b}(k)$ depends only on $k \equiv |\bm k|$ follows from rotational invariance. In the case when $a=b$, this 2-point autocorrelation function defines the power spectrum of the field $P_a(k) \equiv P_{a a}(k)$, which characterises its density fluctuations across different scales $k$.

The transfer function $T_{a b}(k)$ and cross-power spectrum $C_{a b}(k)$ between $\delta_a(\mathbf{x})$ and $\delta_b(\mathbf{x})$ are then defined as:
\begin{equation}
    T_{a b}(k)=\sqrt{\frac{P_a(k)}{P_b(k)}} ; \quad C_{a b}(k)=\frac{P_{a b}(k)}{\sqrt{P_a(k) \times P_b(k)}} \, .
\end{equation}
When the two fields are identical, both of these metrics are equal to $1$ for all $k$, while for two uncorrelated fields we would have $C_{a b}(k) \simeq 0$. Deviations of $T_{a b}(k)$ and $C_{a b}(k)$ from $1$ thus measure the difference in the fields' amplitude and phase, respectively, as a function of scale~$k$.

\section{The Hartley transform}
\label{appendix_hartley}
In this appendix we provide some mathematical details about the Hartley transform, which is closely related to the more familiar Fourier transform \citep{1694454, Bracewell:83}. For a 3D real-valued field $\delta (\mathbf{x})$ defined on a grid (i.e. $\mathbf{x} \in (l,m,n)L/N_{\rm g}$, where $N = N_{\rm g}^3$, we recall that its discrete Fourier transform is given by 
\begin{equation}
    \tilde{\delta} (\bfk) = \bm U \{\delta (\mathbf{x})\}  \equiv  \frac{1}{\sqrt{N}} \sum_{l,m,n} \delta_{l,m,n} e^{-2\pi \mathrm{i} (ul +vm +wn)/N_{\rm g}},
\label{eq:DFT}
\end{equation}
where the integers $(u,v,w)$ label the different modes $\bfk \in (u,v,w) \, 2\pi/L$. The Hartley transform is similarly defined as
\begin{multline}
    \hat{\delta} (\bfk) = \bm H \{\delta (\mathbf{x})\} \equiv \Re(\bm U \{\delta (\mathbf{x})\}) - \Im(\bm U \{\delta (\mathbf{x})\}) \\ 
    = \frac{1}{\sqrt{N}} \sum_{l,m,n} \delta_{l,m,n} \mathrm{cas} \left({2\pi(ul +vm +wn)/N_{\rm g}}\right),
\label{eq:DHT}
\end{multline}
where ${\rm cas}(x) \equiv \cos(x)+\sin(x)$.\footnote{Note that the Fourier transform in \eqref{eq:DFT} can be similarly written in terms of ${\rm cis}(x) \equiv \cos(x) + \mathrm{i} \sin(x) = e^{\mathrm{i} x}$ according to Euler's formula.} From the previous definitions, we see that the Fourier transform maps a real input of shape $(N_{\rm g},N_{\rm g},N_{\rm g})$ to a complex output of shape  $(N_{\rm g},N_{\rm g},N_{\rm g}/2+1)$ (due to the Hermitian condition $\tilde{\delta}(-\bfk) = \tilde{\delta}^{*}(\bfk)$, where we assume that $N_{\rm g}$ is even without loss of generality), while the Hartley transform has the advantage that it maps this real input to a real output of the same shape, i.e. $(N_{\rm g}, N_{\rm g}, N_{\rm g})$. Furthermore, the Hartley transform satisfies all the properties that are required by our formalism, as we proceed to discuss now.

First of all, an inverse operation can also be defined for the Hartley transform, with the added benefit that it is its own inverse, $\bm H^{-1} = \bm H$. Indeed, from Eq. (\ref{eq:DHT}) we can write alternatively $\bm H \{\delta (\mathbf{x})\} = \frac{1+\mathrm{i}}{2} \bm U \{\delta (\mathbf{x})\} + \frac{1-\mathrm{i}}{2} \bm U^{*} \{\delta (\mathbf{x})\}$, from which it is easy to show that 
\begin{equation}
    \bm H \{\bm H\{\delta (\mathbf{x})\}\} =  \delta(\mathbf{x}).   
\end{equation}

Secondly, filtering the modes is just as straightforward with the Hartley transform as it is with the Fourier transform. For both transforms, this involves multiplying the modes with a $k$-dependent real attenuation factor, and since the Hartley transform is just a linear combination of real and imaginary parts of the Fourier transform, each Hartley mode is multiplied by the same factor as the corresponding Fourier mode.

Finally, the power spectrum estimator as it is commonly defined in cosmology can be equivalently written in terms of a Fourier or a Hartley transform. To see this, we recall that the power spectrum estimator is given by 
\begin{equation}
\label{eq:Pk_estimator}
    \widehat{P}(k) = \left(\frac{V}{N} \right) \frac{1}{N_{k}}  \sum_{u,v,w \in k} |\tilde{\delta}(\bfk)|^2,
\end{equation}
where the sum is computed over the $N_k$ modes with modulus in the range $[k,k+\Delta k]$, with $\Delta k = 2\pi/L$. From the definition in Eq. (\ref{eq:DHT}), we have $\Re(\tilde{\delta}(\bfk)) = (\hat{\delta}(\bfk)+\hat{\delta}(-\bfk))/2$ and $\Im(\tilde{\delta}(\bfk)) = (\hat{\delta}(-\bfk)-\hat{\delta}(\bfk))/2$, which implies 
\begin{equation}
    |\tilde{\delta}(\bfk)|^2 = \frac{1}{2} \left( \hat{\delta}^2(\bfk)+\hat{\delta}^2(-\bfk) \right).   
\end{equation}
Inserting the previous identity in Eq. (\ref{eq:Pk_estimator}), and using $|\bfk| = |-\bfk| = k$, we arrive at the following expression
\begin{equation}
    \widehat{P}(k) = \left(\frac{V}{N} \right) \frac{1}{N_{k}}  \sum_{u,v,w \in k} \hat{\delta}^2(\bfk).
\end{equation}

\section{TURNING MAP ESTIMATOR INTO A FAST SAMPLER}
\label{appendix_mse}
\begin{figure}
    \includegraphics[width=\columnwidth]{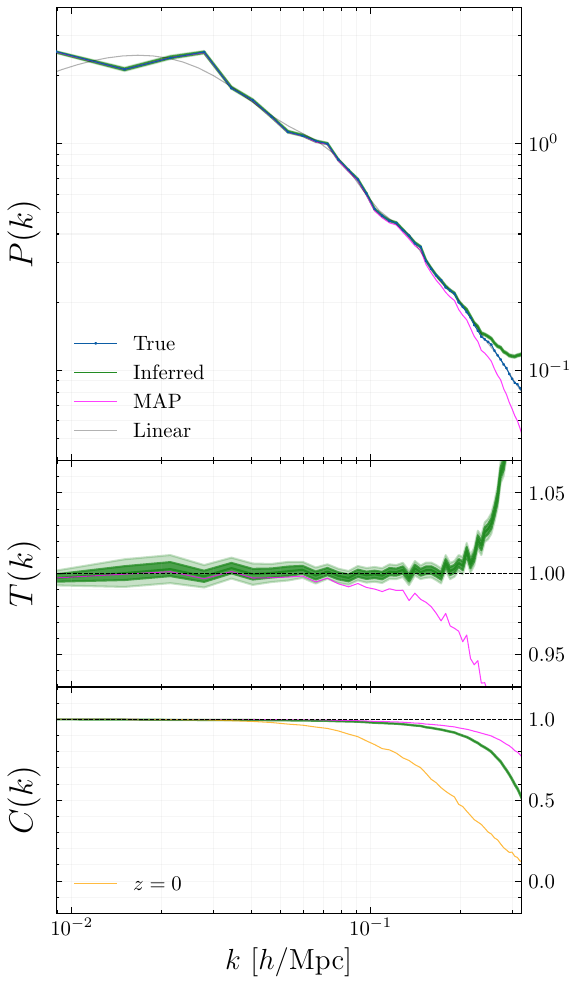}
    \caption{Comparison of the summary statistics at $z=127$ between the ground truth simulation and the samples produced with an MSE-trained MAP estimator and the Gaussian analytical fit for the precision matrix (\ref{eq:fit}), following the algorithm described in Sec. \ref{subsec:fit}. Our generated samples reproduce the true field with a high level of precision up to $k \simeq 0.2 \, h/\text{Mpc}$.}
    \label{fig:mse}
\end{figure}

Here, we show the summary statistic comparison plot analogous to the plot in Fig.\ \ref{fig:sum}, obtained with samples produced following the algorithm described in Sec. \ref{subsec:fit}. First, we train a MAP point estimator (\ref{eq:MAP}) with a simple MSE loss function:
\begin{equation}
    \mathcal{L} = \sum_{i=1}^n \left(\bfz_i - \mathbf{\hat{\boldmu}}_\bftheta(\bfx_i)\right)^2 \, ,
\end{equation}
then apply it to the observed field $\bfx_{\mathrm{obs}}$ and generate 1000 samples via the Gaussian analytical fit for the precision matrix (\ref{eq:fit}). As can be seen from Fig. \ref{fig:mse}, the resulting samples have the same agreement in the power spectrum and cross-correlation up to $k \simeq 0.2 \, h/\text{Mpc}$ as with the fully trained ansatz (\ref{eq:posterior}), thereby validating our method for turning any IC point estimator into a fast IC sampler.

\bsp	
\label{lastpage}
\end{document}